**Global apocalypse at the turn of the first Millennium AD? Climate fluctuations, astronomic phenomena and socio-political turbulences in 10th and 11th century Byzantium and Japan in comparative perspective**

Johannes Preiser-Kapeller, Institute for Medieval Research/Department for Byzantine Research, Austrian Academy of Sciences (Johannes.Preiser-Kapeller@oeaw.ac.at)

**Abstract:** Around the turn of the first Millennium AD, both in Christian polities such as the Byzantine Empire as well as in Buddhist communities such as in Heian Japan, expectations of an end of times emerged. Although based on different religious and independent chronological interpretations, they gained attraction at the same time due to the parallel observation and interpretation of the same astronomical phenomena (such as sightings of Halley´s comet in 989 AD) or of simultaneous climate anomalies, which can partly be connected with the Oort Solar Minimum of the 11th century. This paper explores and compares the interplay between natural phenomena, political and religious unrest and apocalyptic interpretations on the basis of historical and natural scientific evidence.

**Introduction**

The turn of the first Millennium AD among some Christian communities was related to apocalyptic expectation. These fears and hopes clustered not only in the decades before and after the year 1000 AD (anno Domini, years since the birth of Jesus Christ), but also at later dates throughout the 11th century, such as around 1030 AD (one thousand years after the baptism respectively later crucifixion and resurrection of Jesus Christ) or around 1064/1065 AD (based on calculations on a recurrence of the same Easter date as in the year of the resurrection). Such speculations motivated authors in Western Europe, in the Byzantine Empire (such as Leo the Deacon) or in Armenia (such as Matthew of Edessa).[1] Furthermore, in the same period, but based on different chronologies, expectations of a turn of times arose among Islamic communities (such as in some circles of the al-Ismāʿīlīya) as well as in parts of the Buddhist world (connected with ideas of a "Degenerate Age of Dharma", in Japanese *mappō*).[2] To illustrate and support their visions of history and current affairs, authors interpreted celestial signs (such as sightings of Halley's Comet in 989 and 1066), extreme meteorological phenomena (droughts, floods) and other disasters (such as earthquakes or epidemics among humans and animals) as portents of the imminent apocalypse. In fact, modern historical climatology identified an increase of the frequency of extreme events across Afro-Eurasia from Western Europe via the Eastern Mediterranean (Byzantium, the Fatimid Empire), the Middle East and Central Asia to China and Japan, which can be partly related to the "Oort Minimum" of solar activity between 1010 and 1080 CE. Earlier clusters of calamities can be related to climate anomalies in the aftermath of major volcanic eruptions (see below).

---

[1] Landes, "Lest the Millennium Be Fulfilled"; Fried, "Endzeiterwartung um die Jahrtausendwende"; Palmer, *The Apocalypse in the Early Middle Ages*; Magdalino, "The Year 1000 in Byzantium"; Brandes, "Byzantine Predictions of the End of the World".
[2] Cook, "Messianism and Astronomical Events". For Buddhist expectations, see below.

These natural phenomena, however, were not the cause for apocalyptic expectations, but selectively integrated and interpreted by the observers and authors of the time in their texts. Based on individual narrative strategies, more or fewer calamities were reported for specific periods or rulers in more or less detail. The present paper, based on a more wide-ranging research project, due to limitations of space focuses on Byzantium and Japan for the decades before and after 1000 AD. As it demonstrates, extreme climatic events and other celestial portents were related to the quality of rulership, elites or societies at large within frameworks of what has been called "moral meteorology".[3] Early medieval Japan for instance adapted the Chinese ideal of the "Sage king", who "should pay due attention to omens and portents reported to him and alleviate the lot of the poor in time of drought or disaster by remission of their taxes, himself setting examples of frugality and admonishing his courtiers against luxury."[4] Accordingly, until 820 an office existed at the Heian court for observations and interpretations of celestial phenomena and other portents. This practice continued in less centrally organised form afterwards.[5]

**The Medieval Climate Anomaly and periods of climatic fluctuations**

In 1959, Hubert Lamb (1913-1997), a pioneer of historical climatology, introduced the term "Medieval Warm Period" for the time between the 10th and 13th centuries CE, based on his reading of medieval sources and the then limited number of temperature reconstructions for England.[6] The concept was adapted for other regions, including Japan, where scholarship defined a "Nara-Heian-Kamakura Warm Period" between the 8th and 13th century CE.[7]

---

[3] For this term, see Elvin, "Who Was Responsible for the Weather? Moral Meteorology in Late Imperial China".
[4] Ury, "Chinese learning and intellectual life", p. 357. For similar patterns of interpretation in Byzantium see Magdalino, "Astrology"; Magdalino, "The Year 1000 in Byzantium"; Dagron, "Quand la terre tremble…"; Dagron, "Les diseurs d'événements".
[5] Ury, "Chinese learning and intellectual life", p. 357; Grapard, "Religious practices", pp. 547-557. Between 806 and 1073, "as many as 1,686 natural occurrences of an ominous nature were officially recorded: 653 earthquakes, 134 fires, 89 instances of damage to the crops, 91 outbreaks of epidemics, 356 calamitous occurrences of a heavenly nature (volcanic eruptions, comets, eclipses, thunder in clear skies), and 367 ghostly events".

Based on an increasing number of proxy data[8], however, recent scholarship has demonstrated that the "Medieval Warm Period" was neither continuously warm nor "optimal" neither in Europe nor in Byzantium or Japan, not to mention other parts of the globe.[9] Therefore, the term "Medieval Climate Anomaly" (MCA) was introduced. It marks a period of globally higher average temperatures than the preceding "Late Antique Cold Period" and the succeeding "Little Ice Age", but with strong differences in the regional manifestation of this global climate trend, and interrupted by decades of lower average temperatures, as visible from recent temperature reconstructions for Europe for instance (see **Fig. 3**).[10] Such more turbulent climatic dynamics are equally confirmed by the reconstruction of spring temperatures on the basis of the registered date of the start of the cherry blossom in Kyōto (under the name Heian-kyō capital of Japan since 794). It indicates a decline of spring temperatures in this region of Japan from 970 onwards, with a nadir around the year 1015 and a return to more stable temperature conditions by 1045, followed by two other cold periods in the first and the second half of the 12th century (see **Fig. 4**).[11]

One factor in the temporal dynamics of the Medieval Climate Anomaly (and of global climate in general) were fluctuations in the sun´s activity, which influenced the amount of solar irradiation (as essential form of energy input) that reached planet Earth. The Medieval Climate Anomaly included two solar maxima between ca. 920 and 1010/1020 as well as between ca. 1100 and 1200/1250. Reduced solar activity (and therefore cooler global average temperatures), on the contrast, characterised the "Oort Minimum" (ca. 1010 to 1080), while the "Wolf Minimum" (ca. 1280 to 1345) already marked the transition from the Medieval Climate Anomaly to the "Little Ice Age".[12] Large volcanic eruptions had short-term climatic effects.

---

[6] Rohr/Camenisch/Pribyl, "European Middle Ages"; Summerhayes, *Palaeoclimatology*, p. 440; Pfister/Wanner, *Klima und Gesellschaft in Europa*, pp. 22–23.
[7] Sakaguchi, "Warm and cold stages"; Yamada et al., "Late Holocene monsoonal-climate change"; Razjigaeva et al., "Landscape response". On the interplay between climatic change and agriculture in medieval Japan see also von Verschuer, *Rice, Agriculture, and the Food Supply*, pp. 2-4, 10-11, 34-37, 236-237, 241-242.
[8] For an overview on these types of data see Mathez/Smerdon, *Climate Change*, pp. 229–38, Brönnimann/Pfister/White, "Archives of Nature and Archives of Societies"; Pfister/Wanner, *Klima und Gesellschaft in Europa*, pp. 16–20 and 118–31.
[9] Xoplaki et al., "The Medieval Climate Anomaly and Byzantium"; Adamson/Nash, "Climate History of Asia (Excluding China)". For a reconstruction of European summer temperatures see for instance Luterbacher et al., "European summer temperatures".
[10] Diaz et al. "Spatial and Temporal Characteristics"; Rohr/Camenisch/Pribyl, "European Middle Ages"; Summerhayes, *Palaeoclimatology*, pp. 442–48.
[11] Aono/Saito, "Clarifying Springtime Temperature Reconstructions".
[12] Usoskin, "A History of Solar Activity"; Lean, "Estimating Solar Irradiance since 850 CE", pp. 135 and 142; Mathez/Smerdon, *Climate Change*, pp. 180–82; Summerhayes, *Palaeoclimatology*, pp. 455–64; Guiot et al., "Growing Season Temperatures"; Polovodova Asteman/Filipsson/Nordberg, "Tracing winter temperatures";

Eruptive ejections of aerosols caused atmospheric phenomena, which disquieted contemporary observers. These aerosols could contribute to cooler temperatures over several months due to the reduction of solar irradiation, but also created other and regionally diverse climatic effects (decreased as well as increased temperatures, decreases as well as increases of precipitation). Thus, various forms of weather extremes could emerge from the atmospheric perturbations caused by volcanic eruptions. They initiated short term climatic fluctuations also during periods otherwise characterised by higher and more stable temperature conditions as during the above-mentioned solar maxima, such as an eruption described in written sources and also identified due to its chemical signature in ice cores from Greenland for the year 939 (possibly coming from the Eldgjá on Iceland), for 946 (Paektu Mountain in Korea, whose ashes also reached Japan) or a "cluster" of eruptions (maybe in Iceland and Japan) between 1108 and 1110.[13]

The regional effects of solar and volcanic climate forcing depended on their impact on regular climate oscillations between oceans and continents. For weather conditions in western Afro-Eurasia, the Northern Atlantic Oscillation (NAO) plays a decisive role. Its dynamics are measured in an index of the differences in air pressure between the Iceland low and the high over the Azores. A strong difference between these air pressure regions (resulting in a positive NAO-index) usually causes warmer and wetter weather in Western and Central Europe, but drier conditions in the Mediterranean. A low difference on the contrast results in colder and drier weather in Western and Central Europe, but more humid conditions in the Mediterranean. For the more stable periods of the Medieval Climate Anomaly, such as around 950 CE or 1140 CE, a predominantly positive NAO-index was reconstructed, while a weaker NAO has been identified during the Oort Solar Minimum in the mid-11$^{th}$ century.[14]

A further oscillation pattern is the El Niño-Southern Oscillation (ENSO), described as interplay between an area usually characterised by low air pressure and warm water temperatures in the western Pacific (around modern-day Indonesia) and an area of high air pressure and cooler temperatures off the western coast of South America. These "usual" conditions characterise the

---

Cohen/Stanhill, "Changes in the Sun´s radiation", pp. 691–705; Dorman, "Space weather and cosmic ray effects", pp. 713–31; Campbell, *The Great Transition*, pp. 37–38, 50–54.

[13] Chen et al., "Clarifying the distal to proximal tephrochronology"; Razjigaeva et al., "Landscape response"; Sigl, "Timing and Climate Forcing"; Guillet, "Climatic and societal impacts"; Büntgen et al., "Cooling and societal change"; Büntgen, "Prominent role of volcanism"; Mathez/Smerdon, *Climate Change*, pp. 176–80; Summerhayes, *Palaeoclimatology*, pp. 466–68; Stenchikov, "The role of volcanic activity"; Riede, "Doing palaeo-social volcanology"; Campbell, *The Great Transition*, pp. 55–58; Wozniak, *Naturereignisse im frühen Mittelalter*, pp. 315–19; Pfister/Wanner, *Klima und Gesellschaft in Europa*, pp. 180–82. On volcanic eruptions and earthquakes in Japan during this periods see Katsuta et al., "Radiocarbon analysis of tree ring".

[14] Goosse et al., "The medieval climate anomaly in Europe"; Guiot et al., "Growing Season Temperatures"; Polovodova Asteman/Filipsson/Nordberg, "Tracing winter temperatures"; Lüning et al., "Hydroclimate in Africa"; Mathez/Smerdon, *Climate Change*, pp. 91–97; Summerhayes, *Palaeoclimatology*, pp. 437–38, 464–65; Campbell, The Great Transition, pp. 45–48; Pfister/Wanner, *Klima und Gesellschaft in Europa*, pp. 38–40.

"neutral" state of the Southern Oscillation. The "El Niño" state (usually observed around Christmas off the coast of Peru, hence the name) is characterised by cooler than normal conditions in the western Pacific and warmer ones in the eastern Pacific. Its counterpart, "La Niña", is characterised by warmer than normal water temperatures in the western Pacific warmer and cooler ones in the eastern Pacific. These different states of the Southern Oscillation bring about significant changes in the strength of winds and the distribution of precipitation from the ocean towards the continents.[15] Especially during the Oort Minimum of the 11th century, but also before since the 940s, several El Niño- and La Niña-events have been identified, which influenced the monsoon patterns over East Asia, contributing to longer periods of droughts (also reconstructed by W. W. Farris for Japan from written records[16]), but also to higher frequencies of other weather extremes.[17]

**The "glorious" times of Basil II (976-1025) and Fujiwara no Michinaga (995-1028) and the apocalyptic expectations of Leo the Deacon and Genshin**

The reign of Emperor Basil II (976-1025), due to his military successes in Anatolia and on the Balkans (see **Fig. 1**) , in scholarship is often seen as apex of medieval Roman power, especially when compared to the later severe crisis of the Byzantine Empire in the 11th century.[18]

Also, the times of his predecessors were plagued by calamities. During the late reign of Romanos I Lakapenos (920-944), an epidemic among cattle broke out, which affected the empire for the following years until the reign of his namesake and grandson Romanos II (959-963).[19] Tim Newfield has collected parallel reports on epizootics among cattle from Western Europe between 940 and 944 and suggests a possible connection of the outbreak and spread of the disease with climate anomalies in the aftermath of volcanic eruptions in 939/940 such as the one of the Eldgjá on Iceland.[20] In the early reign of Romanos II in October 960, equally a lack of grain and an increase of prices is reported for Constantinople, which the emperor tried to mitigate with the purchase of grain in "west and east".[21] Weather extremes may have contributed to this shortfall; we read about unusual cold and heavy rains during the (ultimately successful) Byzantine expedition against Arab Crete in 960/961.[22] In 963/964, "there was a great famine in Cilicia [which also impeded some of the Byzantine campaigns in the area], and a great many of the people of the Arabs left and fled to Damascus. And there was also a severe famine in Aleppo, and in Harran and in Edessa".[23] Around the same time, in 963 and 964, parts of Italy were affected by famine, while severe floods affected the provinces along the Yellow River in China between 964 and 968, thus illustrating the geographical dimensions of the climate anomalies in these years.[24]

---

[15] Mathez/Smerdon, *Climate Change*, pp. 71–73; Grove/Adamson, *El Niño in World History*.
[16] Farris, *Japan´s Medieval Population*, pp. 38-40; Farris, "Famine, Climate, and Farming", pp. 278-280, 284; Farris, *Daily Life and Demographics*, pp. 65-67
[17] Sakashita et al., "Relationship between early summer precipitation"; Zhang et al., "Modulation of centennial-scale hydroclimate variations"; Zhang et al., "Evidence of ENSO signals"; Xu et al., "Tree-ring oxygen isotope"; Yamada et al., "Late Holocene monsoonal-climate change".
[18] Kaldellis, *Streams of Gold*, p. xxviii.
[19] John Scylitzes, *Synopsis*, Romanos II, 8, ed. Thurn, pp. 251–52; trans. Wortley, pp. 242–43. Schminck, "Zur Einzelgesetzgebung", 281, note 73 (arguing of a dating of the lack of grain to October 961). For further evidence see also PmbZ nr 26834, note 17.
[20] Newfield, "Early Medieval Epizootics"; Newfield, "Domesticates, disease and climate". See also Sigl, "Timing and Climate Forcing"; Wozniak, *Naturereignisse im frühen Mittelalter*, pp. 680–81. For references to climate extremes in the 940s in Egypt and Mesopotamia see also Telelis, *Μετεωρολογικά φαινόμενα*, nr 376 and 377.

Byzantium was perturbed by further portents and catastrophes, at least according to the apocalyptically inspired history of Leo the Deacon (ca. 950-995), who mentions an earthquake in northwestern Asia Minor in 967 and a severe storm and flooding in Constantinople and its environs on 21 June of the same year, so that "people wailed and lamented piteously, fearing that a flood like that fabled one of old was again befalling them."[25] Furthermore, on 22 December 968 "an eclipse of the sun took place", so that once more "people were terrified at the novel and unaccustomed sight, and propitiated the divinity with supplications, as was fitting". As Leo does not forget to mention, he was an eyewitness, since "at that time I myself was living in Byzantium, pursuing my general education."[26]

In addition to these short-term portents and calamities, a shortage of grain of three or even five years duration reportedly troubled the population during the reign of Nikephoros II Phokas (963-969).[27] The causes for this famine are not mentioned, but in the *History* of John Scylitzes we are informed that in May 968 "there were fierce, burning winds (…), which destroyed the crops, even the vines and trees, with the result that in the twelfth year of the indiction there was an intense famine."[28] A change towards more arid conditions in the late 960s, which continued until the early 11th century, is also indicated in the isotope data from the speleothems in the Sofular cave in Northwestern Asia Minor (see **Fig. 5**).[29]

---

[21] Theophanes Continuatus VI, 13, ed. Bekker, p. 479, 1–11. Telelis, Μετεωρολογικά φαινόμενα, nr 394; Teall, "The Grain Supply"; Kaldellis, *Streams of Gold*, p. 32. For references to extreme events and famines in neighbouring regions of Byzantium such as Armenia and Mesopotamia during the 950s see Matthew of Edessa, *History* I, 1, trans. Dostourian, p. 19; al-Maqrīzī, *Ighāthah*, trans. Allouche, p. 29; Bar Hebraeus, *Chronography*, trans. Budge, p. 165 and 167; Telelis, Μετεωρολογικά φαινόμενα, nr 388 and 391–393.
[22] Telelis, Μετεωρολογικά φαινόμενα, nr 395; Kaldellis, *Streams of Gold*, pp. 34–38.
[23] Bar Hebraeus, *Chronography*, trans. Budge, p. 170. Telelis, Μετεωρολογικά φαινόμενα, nr 398; Hassan, "Extreme Nile floods".
[24] Wozniak, *Naturereignisse im frühen Mittelalter*, p. 615 (with citation of sources); Kaldellis, *Streams of Gold*, pp. 46–47; Zhang, *The River, the Plain, and the State*, pp. 110–12; Mostern, *The Yellow River*, pp. 123–25.
[25] Leo the Deacon, *History* IV, 9, ed. Hase, pp. 69–70; trans. Talbot/Sullivan, pp. 117–119. Telelis, Μετεωρολογικά φαινόμενα, nr 402; Wozniak, *Naturereignisse im frühen Mittelalter*, pp. 287–88 (on the earthquakes). The ashes could have been the result of an eruption of Vesuvius in 968, see Wozniak, *Naturereignisse im frühen Mittelalter*, pp. 329–30.
[26] Leo the Deacon, *History* IV, 11, ed. Hase, p. 72; trans. Talbot/Sullivan, pp. 122–23. Telelis, Μετεωρολογικά φαινόμενα, nr 402; Wozniak, *Naturereignisse im frühen Mittelalter*, pp. 196, 216–17.
[27] John Scylitzes, *Synopsis*, John I, 3, ed. Thurn, pp. 286, 48–56; trans. Wortley, pp. 273–74; Leo the Deacon, *History* VI, 8, ed. Hase, p. 103; trans. Talbot/Sullivan, pp. 152–53. Telelis, Μετεωρολογικά φαινόμενα, nr 406.

Basil´s early reign between 976 and 989, however, was equally overshadowed by attempted military coups of two relatives of his predecessors Nikephoros II Phokas and John I Tzimiskes, Bardas Phokas and Bardas Skleros.[30] The succession of two bloody civil wars contributed to the "apocalyptic" mood of the contemporary history of Leo the Deacon, who wrote that "many extraordinary and unusual events have occurred in novel fashion in the course of my lifetime: fearsome sights have appeared in the sky, unbelievable earthquakes have occurred, thunderbolts have struck and torrential rains have poured down, wars have broken out and armies have overrun many parts of the inhabited world, cities and whole regions have moved elsewhere, so that many people believe that life is now undergoing a transformation, and that the expected Second Coming (*deutera katabasis*) of the Saviour and God [Jesus Christ] is near, at the very gates."[31]

For Basil II, this sequence of portents starts in Leo´s text with a comet in August to October 975 (so still during the reign of John I Tzimiskes), which "scholars of astronomy" misinterpreted as sign of future victories, while according to Leo it foretold "bitter revolts, and invasions of foreign peoples, and civil wars, and migrations from cities and the countryside, famines and plagues and terrible earthquakes, indeed almost the total destruction of the Roman empire (…)."[32] Another "sinking" star in August 986 foreboded a defeat of Basil II´s army against the Bulgarians.[33] The sighting of Halley´s Comet between August and September 989, which was equally visible in other parts of Europe and across Asia (including Japan, see below)[34], was followed by further military defeats and especially a devastating earthquake in Constantinople on 25 October 989, which even damaged Hagia Sophia. Furthermore, so Leo, "harsh famines and plagues, droughts and floods and gales of violent winds (…), and the barrenness of the earth and calamities that occurred, all came to pass after the appearance of the star. But my history will describe these in detail in their place."[35] The reference to drought would find a counterpart in the isotope record from the Sofular cave in Northwestern Asia Minor, which indicates the 990s as the driest decade in the entire 10th century (see **Fig. 5**).[36] Tree ring data from modern-day Albania points to very cold conditions in that region in the early 990s.[37] Leo´s history ends, however, shortly after this passage, and the author most probably died at some point before the year 1000. Thus, he did not witness the later military successes of Basil´s II reign, especially his destruction of the Bulgarian Empire in 1018, which earned him the praise of later historians (since the 12th century as "Bulgar Slayer"), who also wrote under the impression of the severe crisis of the empire emerging under Basil´s successors in the 11th century.[38] Therefore, we have few information in Byzantine historiography on the further sequence of "harsh famines and plagues, droughts and floods and gales of violent winds" Leo mentioned as underpinnings of his apocalyptic reading of Basil II´s reign. We find, however, references in other sources from. neighbouring historiographical traditions (such as of the Armenians, see below).

---

[28] John Scylitzes, *Synopsis*, Nikephoros II, 20, ed. Thurn, pp. 277, 37–43; trans. Wortley, p. 266.
[29] Fleitmann et al., "Sofular Cave".
[30] Kaldellis, *Streams of Gold*, pp. 81–102.
[31] Leo the Deacon, *History* I, 1, ed. Hase, p. 4; trans. Talbot/Sullivan, pp. 55–56. On this and other apocalyptic interpretations at this time in Byzantium see Magdalino, "The Year 1000 in Byzantium"; Brandes, "Byzantine Predictions of the End of the World".
[32] Leo the Deacon, *History* X, 6, ed. Hase, p. 169; trans. Talbot/Sullivan, pp. 210–12.
[33] Leo the Deacon, *History* X, 8, ed. Hase, p. 172; trans. Talbot/Sullivan, p. 214; Kaldellis, *Streams of Gold*, pp. 95–96. On this and other observations of this comet see Wozniak, *Naturereignisse im frühen Mittelalter*, pp. 141–42.
[34] Wozniak, *Naturereignisse im frühen Mittelalter*, pp. 106–07.

The most powerful contemporary of Basil II in Japan (see **Fig. 2**) was Fujiwara no Michinaga (966-1028), who since 995 like his forefathers since the mid-9th century de facto ran the affairs of the state, while three emperors, bound to his clan through ties of kinship, took turns on the throne. Michinaga´s regency is considered the peak of power of the Fujiwara family by scholarship, like in the case of Basil II especially when compared with the later decline of its influence over the course of the 11th century.[39]

Similar to the reign of Basil II, however, contemporaneous voices uttered concerns on the conditions of state and society and even saw symptoms of end time. Already in 985, the monk Genshin (942–1017) of the Buddhist Tendai school completed his *Ōjōyōshū* ("Collection of Essential [Passages concerning] Birth [into the Pure Land of Amida Buddha]"). In this text he argued that the world as about to enter a "defiled Latter Age", in which the moral and spiritual capacities of the people declined; this would also become manifest in political and social unrest. His interpretation overlapped with the traditional Buddhist identification of three successive ages of Dharma – the periods of the True, Semblance, and Latter Dharma – of which the last one (in Japanese, *mappō*) would be characterized by a decline of Buddhist teaching and religious practices, impeding the escape from the arduous cycle of reincarnations. A possible way out of this dilemma was proposed by Genshin in the devotion to a saviour buddha called Amida, who resides in a paradise-like "Pure Land" (called the "Land of Supreme Bliss" or *Gokuraku* in Japanese) in the far West. By engaging in various practices, especially the *nenbutsu* ("buddha mindfulness"), and thereby focusing the mind on this buddha, one could be re-born after death in the "Pure Land", which – in contrast to the Japan of the time of Genshin – provided ideal preconditions for spiritual fulfilment and even the achievement of the status of Buddhahood.[40]

---

[35] Leo the Deacon, *History* X, 10, ed. Hase, pp. 175–76; trans. Talbot/Sullivan, pp. 217–18. Kaldellis, *Streams of Gold*, p. 104.
[36] Fleitmann et al., "Sofular Cave".
[37] PAGES 2k Network consortium, Database.
[38] Neville, *Guide to Byzantine Historical Writing*, pp. 124–25; Kaldellis, *Streams of Gold*, pp. 104–05.
[39] Shively/McCullough, "Introduction", pp. 1-2, 4-6; McCullough, "The Heian court", pp. 45-50; Hurst, "Insei", pp. 581-586; Horton, "The influence of the Ōjōyōshū", p. 42; Blair, *Real and Imagined*, pp. 110-113 (also on the religious aspect of this regency).

While signs of the coming of the age of the Latter Dharma had been already identified by the founder of the Tendai school on Mount Hiei, Saichō (767-822), and the concepts of "Pure Land"-Buddhism had been introduced to the school by Saichō´s disciple Ennin (794-864), these ideas gained increased currency during and after the life of Genshin, when the conditions of the time gave further support to the interpretation of imminent *mappō*.[41] Over the course of the 10th century, tensions between the imperial court dominated by the Fujiwara clan and noble leaders in the provinces had increased, competing for access to positions of power and to land (manifest in the proliferation of tax-free estates, the *shōen*). Between 935 and 941, the rebellion of Taira na Masakado in eastern Honshū, who even proclaimed himself "New Emperor" (*shinnō*), shuttered the realm. Around the same time in the year 941 Fujiwara no Sumitomo, leader of pirates in the Inland Sea (see **Fig. 2**), threatened the provision of the capital. Furthermore, minor rebellions of Emishi groups occurred in the provinces of Dewa, Owari and Mino (see **Fig. 2**), while earthquakes, typhoons, fires, floods and epidemics (maybe of smallpox, which had become endemic in Japan since a first major outbreak in 735-737 and contributed to demographic recession until the 11th century) afflicted the capital and its environs. During these troubled years (called by modern Japanese historians the "Discord of the Jōhei [931-938] and Tengyō [938-947] Eras"), starting from 938 the monk Kūya (died in 972) preached Pure Land-teachings in Kyōto.[42]

---

[40] Gukanshō, transl. Brown/Ishida, pp. 423-425; Weinstein, "Aristocratic Buddhism", pp. 510-513; Grapard, "Religious practices", pp. 572-573; Horton, "The influence of the Ōjōyōshū"; Rhodes, "Ōjōyōshū", pp. 253-254; Rhodes, *Genshin's Ōjōyōshū*, pp. 1-6, 129-131; Marra, "The development of mappō thought in Japan. Part 1"; Marra, "The development of mappō thought in Japan. Part 2". For some extracts of the Ōjōyōshū in English translation see Lu, *Japan: A Documentary History*, pp. 121-126.

[41] Groner, *Ryōgen and Mount Hiei*, pp. 2-3; Shively/McCullough, "Introduction", pp. 15-16; Weinstein, "Aristocratic Buddhism", pp. 462-473, 507-508; Adolphson, *The Gates of Power*, pp. 25-26; Rhodes, *Genshin's Ōjōyōshū*, pp. 8-9, 51-55.

[42] Sansom, *A History of Japan to 1334*, pp. 142-146, 244-247; Farris, *Population, Disease and Land*, pp. 53-59; Shively/McCullough, "Introduction", pp. 9-10, 18-19; McCullough, "The Heian court", pp. 30-31, 60-62; Morris, "Land and society", pp. 186-187, 224-235; Kiley, "Provincial administration and land tenure"; Weinstein, "Aristocratic Buddhism", pp. 514-515; Grapard, "Religious practices", pp. 573-574; Takeuchi, "The rise of the

Part of these calamities were attributed to an angry spirit (*goryō*) who took revenge upon those who committed misdeed against them during their lifetimes, especially to the ghost of Sugawara no Michizane. He had risen to high positions at the court under Emperors Uda (887-897) and Daigo (897-930), but succumbed in a power struggle with Fujiwara no Tokihira (871–909) and was exiled to Dazaifu in Kyūshū, where he died in March 903. The death of several members of the imperial family over the next years and decades, the destruction of part of the palace by lightning in 930 (killing four courtiers) and a drought in the same year were attributed to the revengeful ghost of Michizane, who according to the vision of a monk Dōken in 941 claimed responsibility for these calamities. In 947, after further revelations and a severe epidemic, followed by typhoons and droughts, a shrine was erected in Kitano at the western edge of the capital to placate the *goryō*, who eventually became the centre of a cult and finally by an imperial decree of 986 was even acknowledged as "Heavenly Deity".[43]

Extreme events, however, continued over the course of Genshin´s lifetime. In retrospect, the early 12th century *Ōkagami* ("Great Mirror") stated on the reign of Emperor Reizei (967-969): "In various ways everything went well through the reign of Emperor Murakami [946-967[, but at the beginning of the Reizei reign [in 967] everyone felt that the world had somehow come to a time of darkness. And there has been deterioration in worldly affairs ever since."[44] But already before, a drought caused widespread hunger in 957, while a famine in 989 was more confined to Kyōto and its environs. In 974, a smallpox epidemic ravaged the capital; in 976 a strong earthquake and in 980 a storm damaged the capital, including parts of the imperial palace. And in 969, there were rumours of another military rebellion (the "Anna incident").[45] Infights for power in the capital resulted in frequent acts of violence and fires (and a decay of the urban framework especially in the western part of the city).[46]

Factionalism even intensified within Genshin´s Tendai school in its monastic centre of Enraykuji on Mount Hiei near the capital. It had its origins in the conflict between the followers of Ennin (794-864) and those of Enchin (814-891), the fifth *zasu* ("headmaster") of the Tendai school from 868 onwards. While the later faction had dominated on Mt. Hiei during the earlier

---

warriors", pp. 653-664; Lu, *Japan: A Documentary History*, pp. 82-84; Farris, *Japan´s Medieval Population*, pp. 9-10, 26-28; Farris, *Daily Life and Demographics*, pp. 51-53; Rhodes, "Ōjōyōshū", p. 252; Rhodes, *Genshin's Ōjōyōshū*, pp. 60-61, 65-66.
[43] Borgen, *Sugawara no Michizane*, esp. pp. 308-324; Sansom, *A History of Japan to 1334*, pp. 215-216; McCullough, "The Heian court", pp. 58-59; Grapard, "Religious practices", pp. 559-564; Groner, *Ryōgen and Mount Hiei*, pp. 86-87, 205; Farris, *Daily Life and Demographics*, pp. 63-64; Rhodes, *Genshin's Ōjōyōshū*, p. 63, 66-67.
[44] Cited after Gukanshō, transl. Brown/Ishida, p. 381. See also Ōkagami, transl. Craig McCullough, p. 226.
[45] Gukanshō, transl. Brown/Ishida, p. 64; Ōkagami, transl. Craig McCullough, p. 141; McCullough, "The Heian court", pp. 63-64; McCullough, "The capital and its society", pp. 175-176; Farris, "Famine, Climate, and Farming", p. 283; Farris, *Daily Life and Demographics*, pp. 52-53
[46] Lu, *Japan: A Documentary History*, pp. 72-73; McCullough, "The capital and its society", pp. 172-173.

10th century, the influence of the Ennin-faction increased with Genshin´s teacher Ryōgen (912–985), who in 923 joined the community on Mount Hiei and gained the support of the imperial regent Fujiwara no Morosuke (908-960) and his clan. Morosuke´s son Jinzen even became monk and Ryōgen´s disciple. In August 966, Ryōgen was appointed *zasu* and ensured control over the monastic complexes on Mount Hiei within the ranks of the Ennin-faction. For this purpose, he also used a building programme necessitated by a major fire in October 966 – by appointing his followers as supervisors for the many halls rebuilt or newly erected. The monastic community grew to a number of more than 2,700 monks at that time. In order to enforce their will, however, both factions increasingly resorted to violence, employing armed monks (*sōhei*) and even exerting pressure on the government. When in 981 one monk Yokei of the Enchin-faction received a prestigious position as abbot, 160 monks of Ryōgen´s following marched to the residence of chancellor Fujiwara no Yoritada in the capital and in a "forceful protest" (*gōso*) threatened violence until Yokei was forced to resign.[47]

Deterred by these skirmishes, Genshin withdrew from active involvement in the politics of the Tendai school to the Yokawa monastery in 980/981, where he finished his *Ōjōyōshū* in 985. He was not the only contemporary interpreting these events as signs of imminent *mappō* and therefore propagating Pure Land-Buddhism. In his *Amida shinjūgi* ("Ten New Doubts concerning Amida") the monk Zenyu (913-990) claimed: "Because we cannot attain realization quickly in this *sahā* world (which is in) the age of the five defilements, we should seek (birth in) the Land of Supreme Bliss. (…) Now, sentient beings of the present *sahā* world in the age of the five defilements are (characterized by) greed, anger, folly, arrogance, mistaken views, and flattery, and their hearts are lacking in sincerity. Therefore, even if we arouse the aspiration for enlightenment and strive wholeheartedly (for Buddhahood), feelings of envy and slander arise in us and we find it hard to accept (the buddha's teachings) in faith. Among the worlds of the ten directions, (our world) is full of numerous pollutions and evils and is (for this reason) called the land in which it is difficult to teach sentient beings. Moreover, the True Dharma has rapidly perished and there is no hope of realizing the fruit (of Buddhahood)."[48]

Further "pollutions and evils" over the next decades seemed to confirm the interpretation of Zenyu and Genshin, who would then propose more concrete dates for the onset of the age of Latter Dharma (see below).

---

[47] Sansom, *A History of Japan to 1334*, pp. 221-223; Groner, *Ryōgen and Mount Hiei*, pp. 20-43, 66-70, 75-76, 118-125, 167-189, 194-196, 219-220, 229-233; Weinstein, "Aristocratic Buddhism", pp. 486-489; Adolphson, *The Gates of Power*, pp. 4-5, 39-45, 63-65; Rhodes, *Genshin's Ōjōyōshū*, pp. 8-9, 84-85, 111-112, 120-122. For other monastic centres in the period see Blair, *Real and Imagined*.
[48] Cited after Rhodes, *Genshin's Ōjōyōshū*, p. 78.

**Further portents: the "993 event" and calamities in the 990s**

For the late 980s and early 990s, Leo the Deacon informs us that "other calamities were portended (…) by the fiery pillars that were manifested in the north in the middle of the night and terrified those who saw them; for these portended the capture of Cherson [on the Crimea] by the Tauroscythians [Rus, in 989 or 990] and the occupation of Berrhoia by the Mysians [Bulgarians, in 989]."[49] Similar celestial phenomena for the early 990s are reported from Central Europe and Cairo and have been interpreted as sightings of aurora borealis (northern lights), which rarely shows in such southern latitudes as of Constantinople or Egypt.[50] According to recent findings, the causative intensive influx of charged particles in upper layers of the atmosphere was a sign of a period of increased solar activity, which peaked in a massive solar eruption in 993/994, whose chemical signature (higher concentration of isotopes of Carbon14 and Beryllium) has been identified in many sites across the globe in the last years[51] (as it has been for a similar event in 774/775 first detected in data from Japan by Fusa Miyake and his team[52]). While the celestial portents around the year 990 thus find a firm basis in natural sciences, it is unclear to what extent this intensive influx of solar particles equally affected atmospheric circulations and changed "surface weather" patterns, as Timofei Sukhodolov and his team hypothesised.[53]

In 993, however, Genshin was less disquieted by celestial portents, but by the escalation of factionalism within his Tendai order on Mount Hiei. After Ryōgen had died in 985, his disciple Jinzen from the Fujiwara clan became *zasu*, which led to an uproar among the opposing Enchin faction. The conflict intensified even when Jinzen retired in 989 and died already in 990. In 993 finally, the monks of the Ennin faction burnt down some of the buildings of their opponents and violently drove out all of them from Mount Hiei. The Enchin monks (after destroying some of the complexes of the Ennin followers in return) found refuge in the monastery of Onjōji (southeast of Mount Hiei, facing Lake Biwa), which had been one of their strongholds before. The long dispute within the Tendai order thus resulted in a permanent schism.[54]

---

[49] Leo the Deacon, *History* X, 8, ed. Hase, p. 172; trans. Talbot/Sullivan, p. 214.
[50] Wozniak, *Naturereignisse im frühen Mittelalter*, pp. 266-267.
[51] Mekhaldi et al., "Multiradionuclide evidence for the solar origin"; Büntgen et al., "Tree rings reveal globally coherent signature".
[52] Miyake/Nagaya/Masuda./Nakamura, "A signature of cosmic-ray increase in AD 774–775".
[53] Sukhodolov et al, "Atmospheric impacts of the strongest known solar particle storm",
[54] Sansom, *A History of Japan to 1334*, pp. 221-223; Groner, *Ryōgen and Mount Hiei*, pp. 20-43, 66-70, 75-76, 118-125, 167-189, 194-196, 219-220, 229-233; Weinstein, "Aristocratic Buddhism", pp. 491-492; Adolphson, *The Gates of Power*, pp. 42-43, 64-65, 69-70; Rhodes, *Genshin's Ōjōyōshū*, pp. 8-9, 84-85, 111-112, 120-122.

More generally, the early 13th century *Gukanshō* interpreted the reign of Emperor Ichijō (986-1011) as "transitional time (*histosugime*)" and stated: "Following Ichijō's enthronement in 986 at the age of seven, a comet streaked across the sky in the last third of the 6th month of 989 [Halley´s comet, which also caused concerns in Byzantium in the same year, see above]. The era name was changed to Eiso [989–990 AD] in the 8th month of that year. Then came the incomparable disaster known as the Eiso typhoon. And in the following year the era name was changed to Shōryaku [990–995 AD]. A bitter struggle developed on Mount Hiei in 993 between the followers of Chisho and those of Jikaku, when the Senko Hall was completely burned by the former [see above]. And the years 994 and 995 were marked by a terrible epidemic that caused many deaths in and around the capital. (…) eight high-ranking nobles died in 995, and nothing like that had ever occurred before, or has occurred since."[55]

This epidemic outbreak (from which Fujiwara no Michinaga profited to a certain extent, since several of his rivals for power fell victim to it) was again attributed to the activities of a *goryō*; therefore, various rituals were executed at Mt. Funaoka to the north of the capital to appease the spirit.[56] In 996, a widespread famine affected Japan, which continued on a more regional level around the capital in the year 997 (when it was also affected by fire); in 998 and 1000, Kyōto again suffered from epidemics. In addition, pirates from the Korean peninsula and the island of Amami-Ōshima plundered the coasts of Kyūshū and adjacent areas between 997 and 999.[57]

These events in Byzantium and Japan overlap with information on portents, weather extremes and calamities affecting Western Asia and North Africa in the 990s.[58] The Armenian chronicler Matthew of Edessa mentions that "it happened at the beginning of the year 446 of the Armenian era [997/998] that a certain comet arose in the heavens and it became visible with a horrible and dreadful appearance, bright and marvellous."[59] Around the same time, during the early reign of the Fatimid Caliph al-Ḥākim (996-1021) in Egypt, "a period of inflation occurred (…). It was caused by an insufficient level of the Nile, which reached only sixteen cubits and a few fingers. Prices rose sharply and wheat was in high demand but was unattainable. The populace lived in a heightened state of fear, women were kidnapped in the streets, and the situation deteriorated. The price of bread reached one dirham for four ratls; then the situation eased when prices dropped."[60] For the area of Baghdad, Bar Hebraeus reports for the winter 998/999 that "severe frost took place (…), and thousands of the palm-trees (…) were destroyed. And those which remained only after very many years acquired straightness".[61]

---

[55] Gukanshō, transl. Brown/Ishida, p. 67.
[56] Sansom, *A History of Japan to 1334*, pp. 159-161; Ōkagami, transl. Craig McCullough, p. 39 (Introduction); Rhodes, *Genshin's Ōjōyōshū*, pp. 62-63; Farris, *Daily Life and Demographics*, pp. 52-53.
[57] McCullough, "The Heian court", pp. 94-95; McCullough, "The capital and its society", pp. 145-146; Farris, "Famine, Climate, and Farming", p. 283; Farris, *Daily Life and Demographics*, pp. 52-53.
[58] Some parallel observations exist for Western Europe, see Wozniak, *Naturereignisse im frühen Mittelalter*, pp. 494–96, 617–21.
[59] Matthew of Edessa, *History* I, 41, trans. Dostourian, p. 41.

**The supernova of 1006 and calamities of the first years of the new Millennium**

As mentioned above, apocalyptic expectations continued in the Christian world throughout the 11[th] century also after the year 1000 AD has passed, supported by further portents and climatic extremes. For the year 1006, the Annals of the monastery of St. Gallen (in modern-day Switzerland) report that „a new star of unusual size appeared, brilliant in appearance and dazzling to the eyes, not without horror. In a wonderful way, sometimes it got darker, sometimes it got lighter and sometimes it disappeared. It was visible for three months in the deepest borders of the South, more than any sign that can be seen in the sky."[62] This phenomenon was equally observed by ʿAlī ibn Ridwān (988–1061), who worked as physician in Cairo in Egypt: "Now I will describe for you an event/spectacle [*aṯār*] that I saw at the beginning of my training. This event occurred in Scorpio, opposite the Sun, at the time the Sun was 15 degrees in Taurus and the spectacle was 15 degrees in Scorpio. It was a large *naizak* [comet], round in shape, and its size two and a half or three times the size of the planet Venus. Its light illuminated the horizon and was very sparkling. The magnitude of the brightness was a little over a quarter of the moon's brightness. It continued to appear and move in that sign of the zodiac with the movement of the equator [diurnal rotation] until the sun arrived in the constellation Virgo, one sixth more, it stopped all of a sudden."[63] Modern-day astronomy had identified these descriptions as observations of a supernova eruption, which was visible worldwide in April to May 1006; the remains of this stellar explosion (NGC [New General Catalogue] nr 5882) can still be found in the constellation Lupus, at a distance of ca. 7,700 lightyears from our sun.[64]

Matthew of Edessa refers to a "certain star, appearing in the form of fire", which "during the reign of Basil, the Greek emperor (…) arose in the heavens, an omen of the wrath of God towards all living creatures and also a sign of the end of the world" for the year 452 of the Armenian era (1003-1004 AD). As in other cases of celestial phenomena, his chronology (writing more than a century after the events) may be confused, and he actually may refer to the supernova of 1006. According to Matthew, this portent was followed by "a violent earthquake throughout the whole land, to such an extent that many thought that the day of the end of the world had arrived. Like the time of the flood all living creatures shook and trembled, and many fell down and died from fear of the intensity of this wrath. After this outpouring of God´s wrath a plague (…) came upon the area and spreading through many regions, reached Sebasteia [modern-day Siwas, in Byzantine Cappadocia, where many Armenians lived]. This plague clearly manifested itself on men's bodies and, because of its harshness, many had no time to make their confession or take communion. Men and beast diminished from the land, and those remaining quadrupeds roamed about the countryside without anyone to take care of them."[65]

---

[60] al-Maqrīzī, *Ighāthah*, trans. Allouche, p. 31. Telelis, Μετεωρολογικά φαινόμενα, nr 419–420; Hassan, "Extreme Nile floods".

[61] Bar Hebraeus, *Chronography*, trans. Budge, pp. 181–82. Telelis, Μετεωρολογικά φαινόμενα, nr 421; Busse, *Chalif und Großkönig*, p. 388–89.

[62] Annales Sangallenses maiores, ad a. 1006, cited after Wozniak, *Naturereignisse im frühen Mittelalter*, p. 81.

[63] Cited after Wozniak, *Naturereignisse im frühen Mittelalter*, p. 79.

[64] http://www.wikisky.org/starview?object_type=3&object_id=276&object_name=NGC+5882&locale=EN; Wilson, *Candidates for historical supernovae*, pp. 80-83; Wozniak, *Naturereignisse im frühen Mittelalter*, p. 83.

Beyond Anatolia and Armenia, between 1005 and 1008, again low Nile floods caused shortages of food and a rise of grain and bread prices in Egypt, which Caliph al-Ḥākim and his officials tried to mitigate with price regulations and drastic measures (such as flogging and public parading) against millers, bakers, hoarders of grain and speculators, who were suspected to take advantage of the misery of the population – obviously with some success, since "prices decreased and harm was averted (…)."[66] In 1007, "snow fell in Baghdad", but the next harvest brought "great abundance" and low prices for wheat. But later, "violent black winds" in the area of Tikrīt to the northwest of Baghdad "destroyed many houses and tore up very many palm-trees and olive-trees by the roots; and great ships were sunk in the Sea of Persia".[67] In 1010, "(…) swarms of locusts appeared in the country of Mosul, and the nomads raided the country on all sides, and there was also a great pestilence. And the famine waxed strong in the country of Khorasan [in eastern Iran] until one litre of bread was sold for a gold dinar." People would even resort to cannibalism, Bar Hebraeus and other sources tell us.[68]

Several floods occurred in various province along the Yellow River in the years between 1000 and 1014 in the reign of Emperor Song Zhenzong (997-1022).[69] Accordingly, the astrological specialists were unsure in their interpretation of the celestial phenomena caused by the supernova eruption of 1006, as we read: "During the third year of the Jing-De reign period (1006 AD) a large star appeared at the west of the Di lunar mansion. Nobody could identify its (omen category); some said that it was an 'ominous star' of the Ke Huang type, which forewarned a disastrous war. At that time Zhou Ke-ming was away on duty in the southern part of China. On his return he spoke to the emperor saying that according to the Tian Wen Lu and the Qing-Zhou Chang the star should be identified as Zhou-Bo, which is (supposed to be) yellow in its colour and brilliant in its brightness. As an 'auspicious star', it would bring great prosperity to the state over which it appeared. He had noticed on his way back that people inside and outside the capital were quite confused over the matter. For this reason he asked the emperor to allow all civil and military officers to celebrate the occasion to calm the people. The emperor praised him and followed his suggestion. He then promoted him to the post of Librarian and Escort of the Crown Prince."[70]

---

[65] Matthew of Edessa, *History* I, 46, trans. Dostourian, p. 43.
[66] al-Maqrīzī, *Ighāthah*, trans. Allouche, pp. 31–33. Telelis, Μετεωρολογικά φαινόμενα, nr 424–427, 430–431; Hassan, "Extreme Nile floods"; Wozniak, *Naturereignisse im frühen Mittelalter*, p. 619.
[67] Bar Hebraeus, *Chronography*, trans. Budge, pp. 183–184. Telelis, Μετεωρολογικά φαινόμενα, nr 428–429; Busse, *Chalif und Großkönig*, p. 389.
[68] Bar Hebraeus, *Chronography*, trans. Budge, p. 185. Busse, *Chalif und Großkönig*, p. 389. Another plague of locusts occurred in Baghdad in 1018, see Bar Hebraeus, *Chronography*, trans. Budge, p. 185. In general, on the "topos" of cannibalism during famines see Wozniak, *Naturereignisse im frühen Mittelalter*, pp. 731–39.
[69] Zhang, *The River, the Plain, and the State*, p. 37, 110–12; Mostern, *The Yellow River*, pp. 142–44.

Again, the mood of the time was less optimistic in Japan. In 1008, extreme rainfall damaged the harvest and led to famine around Kyōto.[71] These and other portents may have confirmed Genshin´s expectation of the onset of the Age of Latter Dharma. In one of his last texts (*Ryōzen'in shiki*, "Liturgy of Ryōzen Cloister") Genshin in 1007 presented two calculations on the number of years that had passed since Buddha Śākyamuni had entered *nirvāṇa*. According to the first one, in 1007, 1,963 years had elapsed since that time, while one of Genshin's contemporaries, Xingchan, calculated that it had been 1,990 years. The beginning of the Age of Latter Dharma was expected for the year 2000 after Buddha´s death, therefore *mappō* would begin either in the year 1017 or 1041 AD.[72]

This interpretation again resonates with later ones such as in the late 11th century *Eiga Monogatari* ("Tales of Splendor") which on the occasion of a palace fire of 1013 states: "World affairs became quite disturbed and people died. Actually, the Emperor's feelings were just right, and the Minister was not bad. But it seems that such things [as this disastrous fire] had to occur because the world is in its Final Age (*yo no sue*). Epidemics break out every year and people die. Many things occur that are very disturbing."[73] In the middle of 1016, another great fire destroyed over 500 houses in the capital.[74]

**Conclusion: "Strange Parallels"**

---

[70] History of Song, ch. 461, cited after Wilson, *Candidates for historical supernovae*, pp. 81-82.
[71] Farris, "Famine, Climate, and Farming", p. 298.
[72] Rhodes, *Genshin's Ōjōyōshū*, pp. 120-123, 130-131, 168-173, 178; Gukanshō, transl. Brown/Ishida, pp. 423-424.
[73] Cited after Gukanshō, transl. Brown/Ishida, p. 375.
[74] Ōkagami, transl. Craig McCullough, pp. 355.

It is unclear, which calculation for the onset of Age of Latter Dharma Genshin himself accepted. He lived to see the year 1017 AD, in which he died.[75] As mentioned above, the frequency of climatic fluctuations increased around this time which the incipient Oort Minimum of solar activity. This was also the case for the Byzantine Empire in the years after the death of Emperor Basil II in 1025, especially in the 1030s and 1040s.[76] In addition, 1000th anniversaries of the baptism and later the crucifixion and resurrection of Jesus Christ around 1030 AD, together with portents such as solar eclipses provided further reasons for apocalyptic expectations, as Matthew of Edessa reports: "In 478 of the Armenian reckoning (1029/30), in the years of the Greek Emperor Vasil [Basil II, d. 1025], there appeared in the heavens a frightful and horrible sign, and anger against all creation. On the third of the month of October at the third hour of the day the upper firmament was rent from the east side to the west, the blue sky was split in two and a brilliant light was thrown down on the earth from the north, and the entire earth trembled with a great shaking; and before the light faded there was a shout and a frightful noise over all creation; the sun darkened and the stars appeared as if in the middle of the night, and all the world was clothed in mourning, and all peoples cried out to God with bitter tears. And then after three days all the princes and nobles were assembled by order of the Armenian king Yovhannēs, and they came before the holy *vardapet* Yovhannēs Kozeṙn (…). And when the Armenian princes came to question him and to understand about the marvellous spectacle and sign, they saw that the holy *vardapet* Yovhannēs had fallen upon his face in sorrow and was crying bitterly. And when they questioned him, he gave an answer with a bitter air and miserable sighs and said 'O children, listen to me; woe and wretchedness to all mankind, for behold today is one thousand years since the binding of Satan whom our Lord Jesus Christ bound with his holy cross, and particularly with his holy baptism in the Jordan river. And now Satan has been freed from his bonds, according to the testimony of the vision of John the evangelist, as the angel of God told him that Satan would be bound for 1000 years and would then escape his bonds. And behold today Satan has been freed from his thousand years of bonds, as this is the year 478 in the Armenian era (1029/30). With 552 years gone before, it comes to 1030 years; given thirty years up to Christ's baptism, and there are 1000 up to today. And now because of this the rending of the heavens has occurred."[77] Again, Matthew´s chronology is confused: while Basil II had already died in 1025, a (spectacular) annular solar eclipse was visible in Armenia and across the Mediterranean on 29 June 1033.[78] It was followed by another annular solar eclipse on 18 April 1037, near the date of Easter, which became the occasion for a second apocalyptic prophecy by Yovhannēs Kozeṙn, this time connected with the 1000[th] anniversary of the crucifixion and resurrection of Jesus Christ.[79]

---

[75] Rhodes, *Genshin's Ōjōyōshū*, pp. 120-123, 130-131, 168-173, 178.
[76] Preiser-Kapeller, "A Collapse of the Eastern Mediterranean?".
[77] Translation cited from Andrews, *Mattʿēos Uṙhayecʿi and His Chronicle*, p. 185.
[78] https://eclipse.gsfc.nasa.gov/5MCSEmap/1001-1100/1033-06-29.gif.

In 1017, locust plagued the fields of Japan, and the court ordered the reading of sutras to cast out the insects.[80] Large parts of the country suffered from drought in 1025 and a cold and rainy summer in 1029, both causing hunger, as did a drought around the capital in 1030 (which was answered with rainmaking rituals, performed several times between 1028 and 1044). Supply shortfalls were aggravated by political unrest in these years, such as the rebellion of Taira no Tadatsune between 1028 and 1031; warfare equally led to famine in the affected provinces until 1032. Furthermore, we have hints at regional and over-regional outbreaks of epidemics, contributing to a lack of labour force in various provinces.[81] Equally, the often violent conflicts between the factions within the Tendai order continued and intensified again in the 1030s and 1040s: in 1035, warrior monks from the Enchin monastery of Onjōji attacked the Ennin faction´s complexes on Mount Hiei. Followers of the latter in 1039 protested at the Fujiwara regent´s residence in Kyōto and set in on fire.[82] Expectations of the *mappō* thus found additional food – as did the apocalyptic expectations in Eastern (and Western) Christendom. In contrast to the earlier calculation reported by Genshin in 1007, however, over the course of the 11[th] century and later, the year 1052 (which was preceded by another outbreak of provincial rebellion in 1051, leading to the so-called "Earlier Nine Years´ War") became the most popular date in Japan for the onset of the Latter Dharma.[83]

Victor Lieberman in the second volume of his brilliant comparative study in 2009 identified various "strange parallels" between polities across Afro-Eurasia in the medieval and early modern periods. Among the possible factors leading to "synchronized trajectories", he identified the rhythms of climate change.[84] A combination of the "archives of society" and the "archives of nature" as undertaken in the present paper partly confirms these assumptions, augmented with evidence for other celestial phenomena. Their simultaneous interpretation as portents of end time in two polities at the extreme ends of Asia without any direct contact was also motivated by (coincidentally) overlapping apocalyptic calculations in both Buddhist and Christian traditions. These interpretations found further confirmation in crisis-prone political and socio-economic conditions, such as struggles between the imperial centre and provincial elites for control over agricultural land, in Byzantium as well as in Japan.[85]

---

[79] https://eclipse.gsfc.nasa.gov/5MCSEmap/1001-1100/1037-04-18.gif ; Andrews, *Mattʿēos Uṙhayecʿi and His Chronicle*, pp. 189-197.
[80] von Verschuer, *Rice, Agriculture, and the Food Supply*, pp. 34-37.
[81] Sansom, *A History of Japan to 1334*, pp. 167-169; McCullough, "The Heian court", p. 74; Grapard, "Religious practices", pp. 536-537; Takeuchi, "The rise of the warriors", pp. 664-670; Farris, "Famine, Climate, and Farming", p. 283; Farris, *Daily Life and Demographics*, pp. 52-53.
[82] Weinstein, "Aristocratic Buddhism", pp. 495-496; Adolphson, *The Gates of Power*, pp. 64-65.
[83] Gukanshō, transl. Brown/Ishida, pp. 423-424; McCullough, "The Heian court", pp. 30-31, 74; Grapard, "Religious practices", pp. 572-573; Takeuchi, "The rise of the warriors", pp. 670-675; Rhodes, *Genshin's Ōjōyōshū*, p. 287. On the actual impact of Genshin´s writings during and after his lifetime see also Horton, "The influence of the Ōjōyōshū".
[84] Lieberman, *Strange Parallels: Volume 2*, pp. 398-407.

A comparison between different texts and (independent) historiographies as well as between the "archives of society" and the "archives of nature" may help us to "triangulate" the extent of these climatic fluctuations or the visibility of celestial phenomena across wider distances. As indicated in the introduction, however, he impact of various interpretative frameworks on the narrative strategies of authors at the same time prohibits to use these texts as just another piece of uniform empirical data –to be simply aggregated within long data series in order to determine possible correlations with natural scientific data, as done also in recent studies.[86] This is even true beyond historiography[87] or religious writing for documentary evidence; in 918, for instance, the imperial council of state in Kyōto declared: "When within a district damage occurs from drought, flood, frost, sleet, worms, or locusts, and the chief official makes an exaggerated report, the penalty is seventy blows of the heavy rod."[88] Similar rules were repeated in 927 and later.[89] References to such calamities in texts thus always served a purpose beyond mere reporting, and to identify these motivations as ways of a "social embedding" of climatic or celestial events is at least as relevant for the historians of the environment as the determination of past geo- and biophysical parameters.[90]

---

[85] Kiley, "Provincial administration and land tenure"; Preiser-Kapeller, "Byzantium 1025-1204".
[86] On these issues see Degroot, et al., "Towards a rigorous understanding of societal responses to climate change".
[87] See Krallis, "Historiography as Political Debate".
[88] Cited after Kiley, "Provincial administration and land tenure", pp. 308-309.
[89] Kiley, "Provincial administration and land tenure", pp. 309-310.
[90] Preiser-Kapeller, *Die erste Ernte und der große Hunger*; Preiser-Kapeller, *Der Lange Sommer und die Kleine Eiszeit*.

# Bibliography

**Primary sources**

al-Maqrīzī, *Ighāthah*, trans. A. Allouche, *Mamluk Economics: A Study and Translation of al-Maqrīzī's Ighāthah*, Salt Lake City 1994

Bar Hebraeus, *Chronography*, trans. E.A.W. Budge, *The Chronography of Gregory Abû'l Faraj, the Son of Aaron, the Hebrew Physician, Commonly Known as Bar Hebraeus: Being the First Part of his Political History of the Word*, 2 vol.s, London 1932

Gukanshō, transl. D. M. Brown/I. Ishida, *The Future and the Past. Translation and Study of the Gukansho, an interpretative History of Japan written in 1219*, Berkeley/Los Angeles/London 1979

John Scylitzes, *Synopsis*, ed. I. Thurn, *Ioannis Scylitzae Synopsis historiarum* (Corpus Fontium Historiae Byzantinae 5), Berlin 1973

John Scylitzes, *Synopsis*, trans. J. Wortley, John Skylitzes, *A Synopsis of Byzantine History 811–1057*, with introductions by J.-Cl. Cheynet and B. Flusin and notes by J.-Cl. Cheynet, Cambridge 2010

Leo the Deacon, *History*, ed. Ch. B. Hase, *Leonis diaconi Caloënsis Historiae libri X*, Bonn 1828

Leo the Deacon, *History*, trans. A.-M. Talbot/D.F. Sullivan, *The History of Leo the Deacon. Byzantine Military Expansion in the Tenth Century*, Washington, D.C. 2005

Matthew of Edessa, *History*, trans. A.E. Dostourian, *Armenia and the Crusades, Tenth to Twelfth Centuries. The Chronicle of Matthew of Edessa*, Lanham 1993

Ōkagami, transl. H. Craig McCullough, *Ōkagami. The Great Mirror. Fujiwara Michinaga (966-1027) and His Times*, Princeton, New Jersey 1980

Theophanes Continuatus VI, ed. I. Bekker, *Theophanes continuatus, Joannes Cameniata, Symeon Magister, Georgius Monachus*, Bonn 1838

**Secondary Literature**

Adamson, G. C. D./Nash, D., "Climate History of Asia (Excluding China)", in S. White/Ch. Pfister/F. Mauelshagen (eds.), *The Palgrave Handbook of Climate History*, London 2018, pp. 203–11

Adolphson, M. S., *The Gates of Power. Monks, Courtiers, and Warriors in Premodern Japan*, Honolulu 2000

Andrews, T.L., *Mattʿēos Uṙhayecʿi and His Chronicle. History as Apocalypse in a Crossroads of Cultures*, Leiden 2017

Aono, Y./Saito, Sh., "Clarifying Springtime Temperature Reconstructions of the Medieval Period by gap-filling the Cherry Blossom phenological Data Series at Kyoto, Japan", *International Journal of Biometeorology* 54 (2010), 211–219.

Bianquis, Th., "Une crise frumentaire dans l'Égypte Fatimide*", Journal of the Economic and Social History of the Orient* 23 (1980), 67–101

Blair, H., *Real and Imagined. The Peak of Gold in Heian Japan*, Cambridge, Mass/London 2015

Borgen, R., *Sugawara no Michizane and the Early Heian Court* (Harvard East Asian Monographs 120), Cambridge, Mass./London 1986

Brandes, W., "Byzantine Predictions of the End of the World in 500, 1000, and 1492 AD", in H.-C. Lehner (ed.), *The End(s) of Time(s). Apocalypticism, Messianism, and Utopianism through the Ages*, Leiden 2021, pp. 32–63

Brönnimann, St./Pfister, Ch./White, S., "Archives of Nature and Archives of Societies", in S. White/Ch. Pfister/F. Mauelshagen (eds.), *The Palgrave Handbook of Climate History*, London 2018, pp. 27–36

Büntgen, U., et al., "Cooling and societal change during the Late Antique Little Ice Age from 536 to around 660 AD", *Nature Geoscience* 9 (2016), 231–236, https://doi.org/10.1038/ngeo2652

Büntgen, U., et al., "Tree rings reveal globally coherent signature of cosmogenic radiocarbon events in 774 and 993 CE", *Nature Communication* 9 (2018) 3605, online: https://www.nature.com/articles/s41467-018-06036-0

Büntgen, U., et al., "Prominent role of volcanism in Common Era climate variability and human history", *Dendrochronologia* 64 (2020), online: https://doi.org/10.1016/j.dendro.2020.125757

Busse, H., *Chalif und Großkönig. Die Buyiden im Irak (945-1055)*, Beirut 2004

Campbell, B.M.S., *The Great Transition. Climate, Disease and Society in the Late-Medieval World*, Cambridge 2016

Chen, X.-Y, et al., "Clarifying the distal to proximal tephrochronology of the Millennium (B-Tm) eruption, Changbaishan Volcano, northeast China", *Quaternary Geochronology* 33 (2016), 61–75

Cohen, Sh./Stanhill, G., "Changes in the Sun´s radiation: the role of wisespread surface solar radiation trends in climate change: dimming and brightening", in T.M. Letcher (ed.), *Climate Change. Observed Impacts on Planet Earth*, Amsterdam 2021, pp. 687–709

Cook, D., "Messianism and Astronomical Events during the First Four Centuries of Islam", *Revue des mondes musulmans et de la Méditerranée* 91-94 (2000), 29-52, online: https://journals.openedition.org/remmm/247#tocto1n6

Dagron, D., "Quand la terre tremble…", in G. Dagron, *Idées byzantines*, vol. I, Paris 2012, pp. 3–22

Dagron, G. "Les diseurs d'événements: Approche de l'astrologie orientale", in G. Dagron, *Idées byzantines*, vol. I, Paris 2012, pp. 23–52

Degroot, D., et al., "Towards a rigorous understanding of societal responses to climate change", *Nature* 591 (2021), 539–550

Diaz, H.F., et al., "Spatial and Temporal Characteristics of Climate in Medieval Times Revisited", *Bulletin of the American Meteorological Society* 92/11 (2011), 1487–1500

Dorman, L.I., "Space weather and cosmic ray effects", in T.M. Letcher (ed.), *Climate Change. Observed Impacts on Planet Earth*, Amsterdam 2021, pp. 711–68

Elvin, M., "Who Was Responsible for the Weather? Moral Meteorology in Late Imperial China", *Osiris 13, Beyond Joseph Needham: Science, Technology, and Medicine in East and Southeast Asi*a (1998), 213–37

Farris, W. W., *Population, Disease and Land in Early Japan, 645-900*, Cambridge, Mass./London 1985

Farris, W. W., *Japan´s Medieval Population. Famine, Fertility and Warfare in a Transformative Age*, Honolulu 2006

Farris, W. W., "Famine, Climate, and Farming in Japan, 670–1100", in M. Adolphson/E. Kamens/St. Matsumoto (eds.), *Heian Japan, centers and peripheries*, Honolulu 2007, pp. 275–304.

Farris, W. W., *Daily Life and Demographics in Ancient Japan*, Ann Arbor 2009

Fleitmann, D., et al., "Sofular Cave, Turkey 50KYr Stalagmite Stable Isotope Data", *IGBP PAGES/World Data Center for Paleoclimatology Data Contribution Series* # 2009-132, online: https://www.ncei.noaa.gov/access/paleo-search/study/8637

Fried, F., "Endzeiterwartung um die Jahrtausendwende", *Deutsches Archiv für Erforschung des Mittelalters* 45 (1989), 381–473

Goosse, H., et al., "The medieval climate anomaly in Europe: Comparison of the summer and annual mean signals in two reconstructions and in simulations with data assimilation", *Global and Planetary Change* 84-85 (2012), 35–47

Grapard, A. G., "Religious practices", in D. H. Shively/W. H. McCullough (eds.), *The Cambridge History of Japan, Vol. 2: Heian Japan*, Cambridge 1999, pp. 517–575

Groner, P., *Ryōgen and Mount Hiei. Japanese Tendai in the Tenth Century* (Kuroda Institute Studies in East Asian Buddhism 15), Honolulu 2002

Grove, R./Adamson, G., *El Niño in World History*, London 2018

Guillet, S., et al., "Climatic and societal impacts of a "forgotten" cluster of volcanic eruptions in 1108–1110 CE", *Nature Scientific Reports* 10 (2020), online: https://doi.org/10.1038/s41598-020-63339-3

Guiot, J., et al., "Growing Season Temperatures in Europe and Climate Forcings Over the Past 1400 Years", *PLoS ONE* 5(4) (2010), e9972. doi:10.1371/journal.pone.0009972

Hassan, F.A., "Extreme Nile floods and famines in Medieval Egypt (AD 930-1500) and their climatic implications", *Quaternary International* 173-174 (2007), 101–12

Horton, S., "The influence of the Ōjōyōshū in Late Tenth- and Early Eleventh Century Japan", *Japanese Journal of Religious Studies* 31/1 (2004), 29–54

Hurst, G. C., III, "Insei", in D. H. Shively/W. H. McCullough (eds.), *The Cambridge History of Japan, Vol. 2: Heian Japan*, Cambridge 1999, pp. 576–642

Kaldellis, A., *Streams of Gold, Rivers of Blood. The Rise and Fall of Byzantium, 955 A.D. to the First Crusade*, Oxford 2017

Katsuta, N., et al., "Radiocarbon analysis of tree ring for a catastrophic collapse in the northern Yatsugatake volcanoes: Its implication for seismotectonics in southwest Japan", *Quaternary International* 604 (2021), 68–74

Kiley, C. J., "Provincial administration and land tenure in early Heian", in D. H. Shively/W. H. McCullough (eds.), *The Cambridge History of Japan, Vol. 2: Heian Japan*, Cambridge 1999, pp. 236–340

Krallis, D., "Historiography as Political Debate", in A. Kaldellis/N. Siniossoglou (eds.), *The Cambridge Intellectual History of Byzantium*, Cambridge 2017, pp. 599–614

Landes, R., "Lest the Millennium Be Fulfilled: Apocalyptic Expectations and the Pattern of Western Chronography 100-800 CE", in W. Verbeke/D. Verhelst/A. Welkenhuysen (eds), *The Use and Abuse of Eschatology in the Middle Ages*, Leuven 1988, pp. 137–211

Lean, J.L., "Estimating Solar Irradiance Since 850 CE", *Earth and Space Science* 5 (2018), 133–49

Lieberman, V., *Strange Parallels: Volume 2, Mainland Mirrors: Europe, Japan, China, South Asia, and the Islands: Southeast Asia in Global Context, c.800–1830*, Cambridge 2009

Lu, D. J., *Japan: A Documentary History, Vol. 1: The Dawn of History to the Late Tokugawa Period*, London 2005

Lüning, S., et al., "Hydroclimate in Africa during the Medieval Climate Anomaly", *Palaeogeography, Palaeoclimatology, Palaeoecology* 495 (2018), 309–22

Luterbacher, J., et al., "European summer temperatures since Roman times", *Environmental Research Letters* 11 (2016) 024001, online: https://iopscience.iop.org/article/10.1088/1748-9326/11/2/024001

Magdalino, P., "Astrology", in A. Kaldellis/N. Siniossoglou (eds.), *The Cambridge Intellectual History of Byzantium*, Cambridge 2017, pp. 198–214

Magdalino, P., "The Year 1000 in Byzantium", in P. Magdalino (ed.), *Byzantium in the Year 1000*, Leiden/Boston 2003, 233–270

Marra, M., "The development of mappō thought in Japan. Part 1", *Japanese Journal of Religious Studies* 15/1 (1988), 25–54

Marra, M., "The development of mappō thought in Japan. Part 2", *Japanese Journal of Religious Studies* 15/4 (1988), 287–305

Mathez, E.A./Smerdon, J. E., *Climate Change. The Science of Global Warming and our Energy Future*, New York 2018

Kiley, C. J., "Provincial administration and land tenure in early Heian", in D. H. Shively/W. H. McCullough (eds.), *The Cambridge History of Japan, Vol. 2: Heian Japan*, Cambridge 1999, pp. 236–340

Krallis, D., "Historiography as Political Debate", in A. Kaldellis/N. Siniossoglou (eds.), *The Cambridge Intellectual History of Byzantium*, Cambridge 2017, pp. 599–614

Landes, R., "Lest the Millennium Be Fulfilled: Apocalyptic Expectations and the Pattern of Western Chronography 100-800 CE", in W. Verbeke/D. Verhelst/A. Welkenhuysen (eds), *The Use and Abuse of Eschatology in the Middle Ages*, Leuven 1988, pp. 137–211

Lean, J.L., "Estimating Solar Irradiance Since 850 CE", *Earth and Space Science* 5 (2018), 133–49

Lieberman, V., *Strange Parallels: Volume 2, Mainland Mirrors: Europe, Japan, China, South Asia, and the Islands: Southeast Asia in Global Context, c.800–1830*, Cambridge 2009

Lu, D. J., *Japan: A Documentary History, Vol. 1: The Dawn of History to the Late Tokugawa Period*, London 2005

Lüning, S., et al., "Hydroclimate in Africa during the Medieval Climate Anomaly", *Palaeogeography, Palaeoclimatology, Palaeoecology* 495 (2018), 309–22

Luterbacher, J., et al., "European summer temperatures since Roman times", *Environmental Research Letters* 11 (2016) 024001, online: https://iopscience.iop.org/article/10.1088/1748-9326/11/2/024001

Magdalino, P., "Astrology", in A. Kaldellis/N. Siniossoglou (eds.), *The Cambridge Intellectual History of Byzantium*, Cambridge 2017, pp. 198–214

Magdalino, P., "The Year 1000 in Byzantium", in P. Magdalino (ed.), *Byzantium in the Year 1000*, Leiden/Boston 2003, 233–270

Marra, M., "The development of mappō thought in Japan. Part 1", *Japanese Journal of Religious Studies* 15/1 (1988), 25–54

Marra, M., "The development of mappō thought in Japan. Part 2", *Japanese Journal of Religious Studies* 15/4 (1988), 287–305

Mathez, E.A./Smerdon, J. E., *Climate Change. The Science of Global Warming and our Energy Future*, New York 2018

McCullough, W. H., "The Heian court, 794-1070", in D. H. Shively/W. H. McCullough (eds.), *The Cambridge History of Japan, Vol. 2: Heian Japan*, Cambridge 1999, pp. 20–96

McCullough, W. H., "The capital and its society", in D. H. Shively/W. H. McCullough (eds.), *The Cambridge History of Japan, Vol. 2: Heian Japan*, Cambridge 1999, pp. 97–182

Mekhaldi, F. et al., "Multiradionuclide evidence for the solar origin of the cosmic-ray events of AD 774/5 and 993/4". *Nature Communications* 6 (2015) 8611, online; https://doi.org/10.1038/ncomms9611

Miyake, F./Nagaya, K./Masuda, K./Nakamura, T., "A signature of cosmic-ray increase in AD 774–775 from tree rings in Japan", *Nature* 486 (2012), 240–242

Morris, D., "Land and society", in D. H. Shively/W. H. McCullough (eds.), *The Cambridge History of Japan, Vol. 2: Heian Japan*, Cambridge 1999, pp. 183–235

Mostern, R., *The Yellow River. A Natural and Unnatural History*, New Haven/London 2021

Neville, L., *Guide to Byzantine Historical Writing*, Cambridge 2018

Newfield, T., "Early Medieval Epizootics and Landscapes of Disease: The Origins and Triggers of European Livestock Pestilences", in: S. Kleingärtner/T.P. Newfield/S. Rossignol/D. Wehner (eds.), *Landscapes and Societies in Medieval Europe East of the Elbe. Interactions Between Environmental Settings and Cultural Transformations*, Toronto 2013, pp. 73–113

Newfield, T., "Domesticates, disease and climate in early post-classical Europe: the cattle plague of c.940 and its environmental context", *Post-Classical Archaeologies* 5 (2015), 95–126

PAGES 2k Network consortium, Database S1 - 11 April 2013 version: http://www.pages-igbp.org/workinggroups/2k-network

Palmer, J. T., *The Apocalypse in the Early Middle Ages*, Cambridge 2014

Pfister, Ch./Wanner, H., *Klima und Gesellschaft in Europa. Die letzten tausend Jahre*, Bern 2021

PmbZ: Lilie, R. J., et al., *Prosopographie der mittelbyzantinischen Zeit Online*, Berlin 1998–2013, online: https://doi.org/10.1515/pmbz

Polovodova Asteman, I./Filipsson, H.L./Nordberg, K., "Tracing winter temperatures over the last two millennia using a north-east Atlantic coastal record", *Climate of the Past* 14 (2018), 1097–1118

Preiser-Kapeller, J., "A Collapse of the Eastern Mediterranean? New results and theories on the interplay between climate and societies in Byzantium and the Near East, ca. 1000–1200 AD", *Jahrbuch der Österreichischen Byzantinistik* 65 (2015), 195–242

Preiser-Kapeller, J., "Byzantium 1025-1204", in F. Daim (ed.), *History and Culture of Byzantium* (Brill´s New Pauly 11), Leiden 2019, pp. 59–89

Preiser-Kapeller, J., *Die erste Ernte und der große Hunger. Klima, Pandemien und der Wandel der Alten Welt bis 500 n. Chr.*, Vienna 2021

Preiser-Kapeller, J., *Der Lange Sommer und die Kleine Eiszeit. Klima, Pandemien und der Wandel der Alten Welt von 500 bis 1500 n. Chr.*, Vienna 2021

Razjigaeva, N. G., et al., "Landscape response to the Medieval Warm Period in the South Russian Far East", *Quaternary International* 519 (2019), 215–231

Rhodes, R. F., "Ōjōyōshū, Nihon Ōjō Gokuraku-ki, and the Construction of Pure Land Discourse in Heian Japan", *Japanese Journal of Religious Studies* 34/2 (2007), 249–70

Rhodes, R. F., *Genshin's Ōjōyōshū and the Construction of Pure Land Discourse in Heian Japan*, Honolulu 2017

Riede, F., "Doing palaeo-social volcanology: Developing a framework for systematically investigating the impacts of past volcanic eruptions on human societies using archaeological datasets", *Quaternary International* 499 (2019), 266–77

Rohr, Ch./Camenisch, Ch./Pribyl, K., "European Middle Ages", in S. White/Ch. Pfister/F. Mauelshagen (eds.), *The Palgrave Handbook of Climate History*, London 2018, pp. 247–63

Sakaguchi, Y., "Warm and cold stages in the past 7600 years in Japan and their global sea level changes and the ancient Japanese history", *Bulletin of Department of Geography, University of Tokyo* 15 (1983), 1–31 (in Japanese with English abstract)

Sakashita, W., et al., "Relationship between early summer precipitation in Japan and the El Niño-Southern and Pacific Decadal Oscillations over the past 400 years", *Quaternary International* 397 (2016), 300–306

Sansom, G., *A History of Japan to 1334*, Stanford 1958

Schminck, A., "Zur Einzelgesetzgebung der „makedonischen" Kaiser", *Fontes Minores* 11 (2005), 269–323

Shively, D. H./McCullough, W. H., "Introduction", in D. H. Shively/W. H. McCullough (eds.), *The Cambridge History of Japan, Vol. 2: Heian Japan*, Cambridge 1999, pp. 1–18

Sigl, M., et al., "Timing and Climate Forcing of Volcanic Eruptions for the Past 2,500 Years", *Nature* 523 (2015), 543–49

Stenchikov, G., "The role of volcanic activity in climate and global changes", in T.M. Letcher (ed.), *Climate Change. Observed Impacts on Planet Earth*, Amsterdam 2021, pp. 607–43

Sukhodolov, T. et al, "Atmospheric impacts of the strongest known solar particle storm of 775 AD", *Science Report* 7 (2017) 45257, online: https://doi.org/10.1038/srep45257

Summerhayes, C.P., *Palaeoclimatology. From Snowball Earth to the Anthropocene*, Chichester 2020

Takeuchi, R., "The rise of the warriors", in D. H. Shively/W. H. McCullough (eds.), *The Cambridge History of Japan, Vol. 2: Heian Japan*, Cambridge 1999, pp. 644–710

Teall, J.L., "The Grain Supply of the Byzantine Empire, 330–1025", *Dumbarton Oaks Papers* 13 (1959), 87–139

Telelis, I.G., *Μετεωρολογικά φαινόμενα και κλίμα στο Βυζάντιο*, 2 vols., Athens 2004

Ury, M., "Chinese learning and intellectual life", in D. H. Shively/W. H. McCullough (eds.), *The Cambridge History of Japan, Vol. 2: Heian Japan*, Cambridge 1999, pp. 341–388

Usoskin, I.G., "A History of Solar Activity over Millennia", *Living Reviews in Solar Physics* 10 (2013), 1, online: http://www.livingreviews.org/lrsp-2013-1

von Verschuer, Ch., *Rice, Agriculture, and the Food Supply in Premodern Japan*, London/New York 2016.

Weinstein, St., "Aristocratic Buddhism", in D. H. Shively/W. H. McCullough (eds.), *The Cambridge History of Japan, Vol. 2: Heian Japan*, Cambridge 1999, pp. 449–516

Wilson, G. Ph., *Candidates for historical supernovae and their comparison against known Chinese records*, MA-Thesis, Durham University, online: http://etheses.dur.ac.uk/3130/

Wozniak, Th., *Naturereignisse im frühen Mittelalter. Das Zeugnis der Geschichtsschreibung vom 6. bis 11. Jahrhundert*, Berlin 2020

Xoplaki, E., et al., "The Medieval Climate Anomaly and Byzantium: A review of the evidence on climatic fluctuations, economic performance and societal change", *Quaternary Science Reviews* 136 (2016), 229–52

Xu, Ch., et al., "Tree-ring oxygen isotope across monsoon Asia: Common signal and local influence", *Quaternary Science Reviews* 269 (2021), online: https://doi.org/10.1016/j.quascirev.2021.107156

Yamada, K., et al., "Late Holocene monsoonal-climate change inferred from Lakes Ni-no-Megata and San-no-Megata, northeastern Japan", *Quaternary International* 220 (2010), 122–132

Zhang, L., *The River, the Plain, and the State: An Environmental Drama in Northern Song China, 1048–1128*, Cambridge 2016

Zhang, J., et al., "Modulation of centennial-scale hydroclimate variations in the middle Yangtze River Valley by the East Asian-Pacific pattern and ENSO over the past two millennia", *Earth and Planetary Science Letters* 576 (2021), online: https://doi.org/10.1016/j.epsl.2021.117220

Zhang, Zh., et al., "Evidence of ENSO signals in a stalagmite-based Asian monsoon record during the medieval warm period", *Palaeogeography, Palaeoclimatology, Palaeoecology* 584 (2021), online: https://doi.org/10.1016/j.palaeo.2021.110714

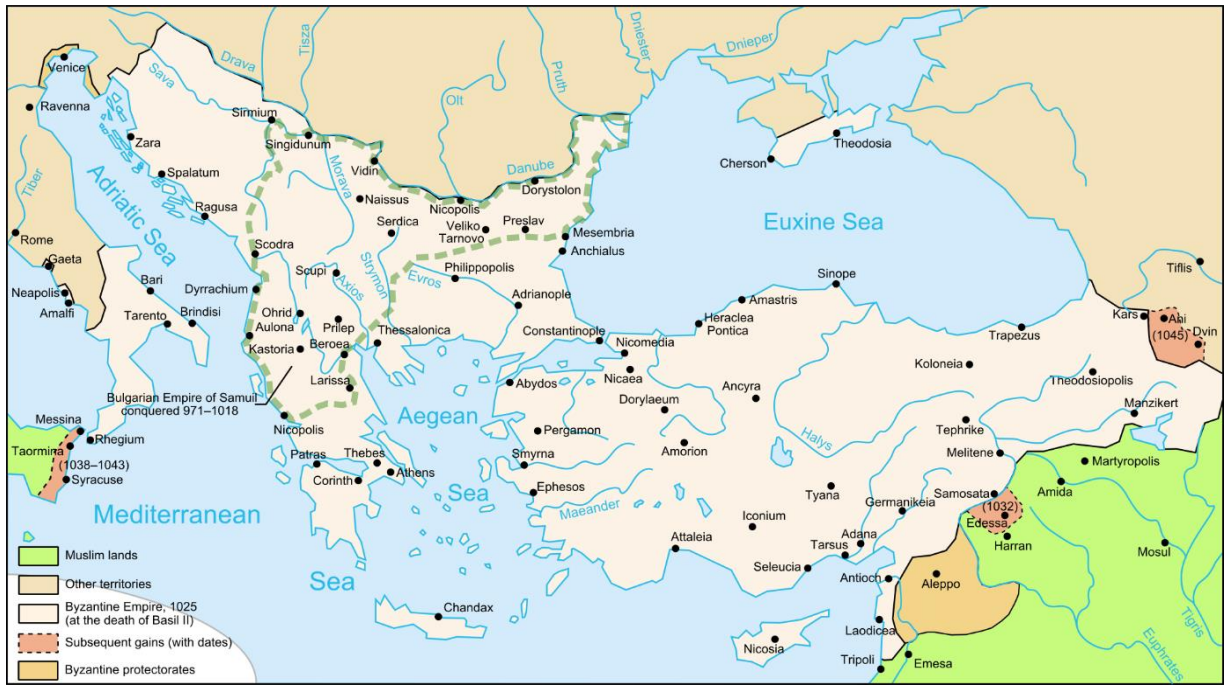

Fig. 1: The Byzantine Empire at the death of Emperor Basil II in 1025 AD
(https://de.m.wikipedia.org/wiki/Datei:Map_Byzantine_Empire_1025-en.svg)

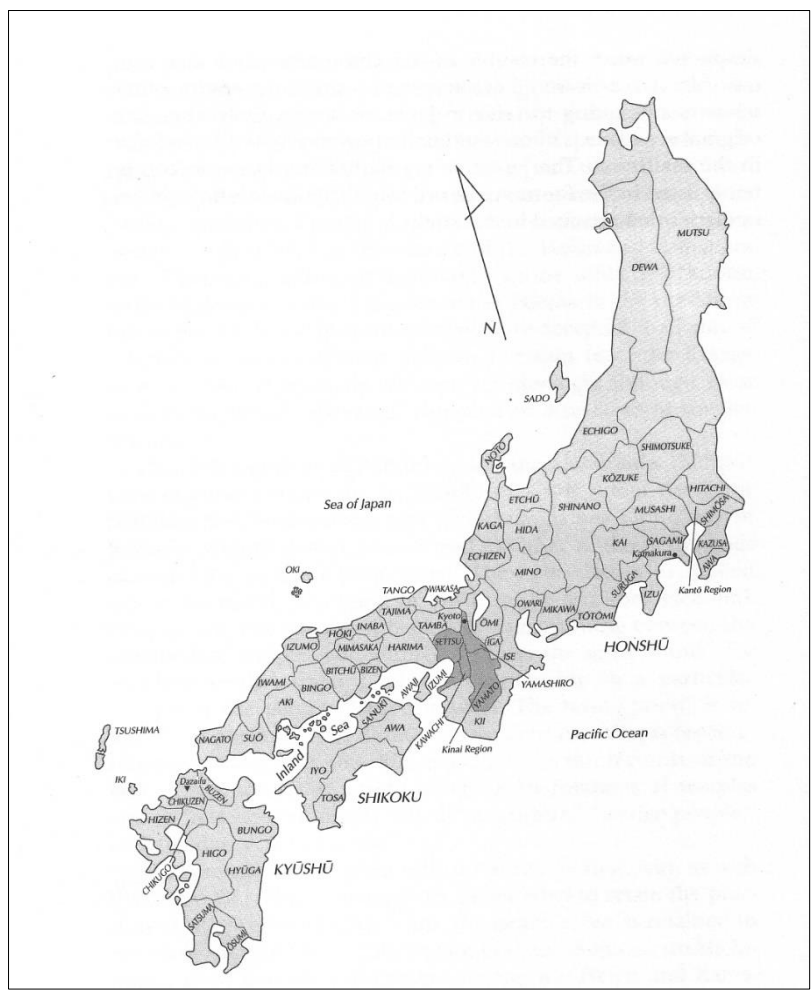

Fig. 2: Japan in the Heian period
(https://sites.fas.harvard.edu/~chgis/japan/images/adolphson_japan.jpg)

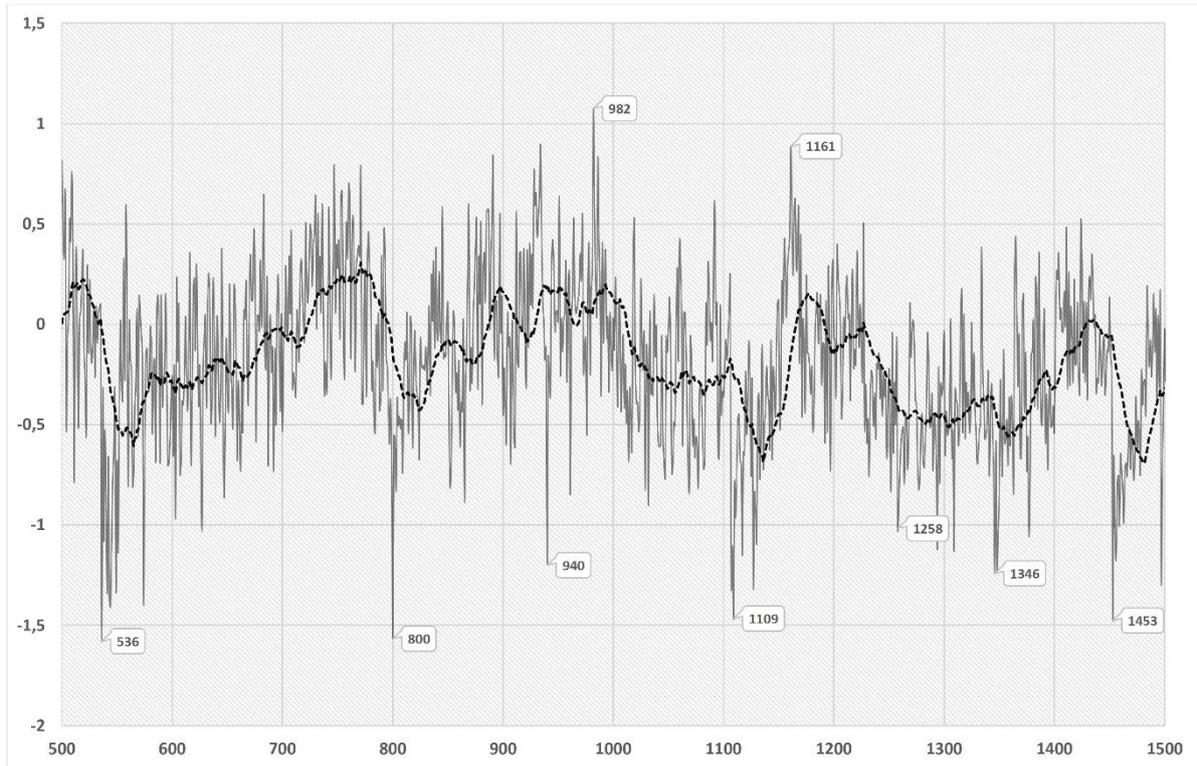

Fig. 3: Average summer temperatures in Western and Central Europe 500–1500 AD, reconstructed on the basis of tree rings (data: Luterbacher et al., "European summer temperatures"; graph: J. Preiser-Kapeller, OEAW, 2022)

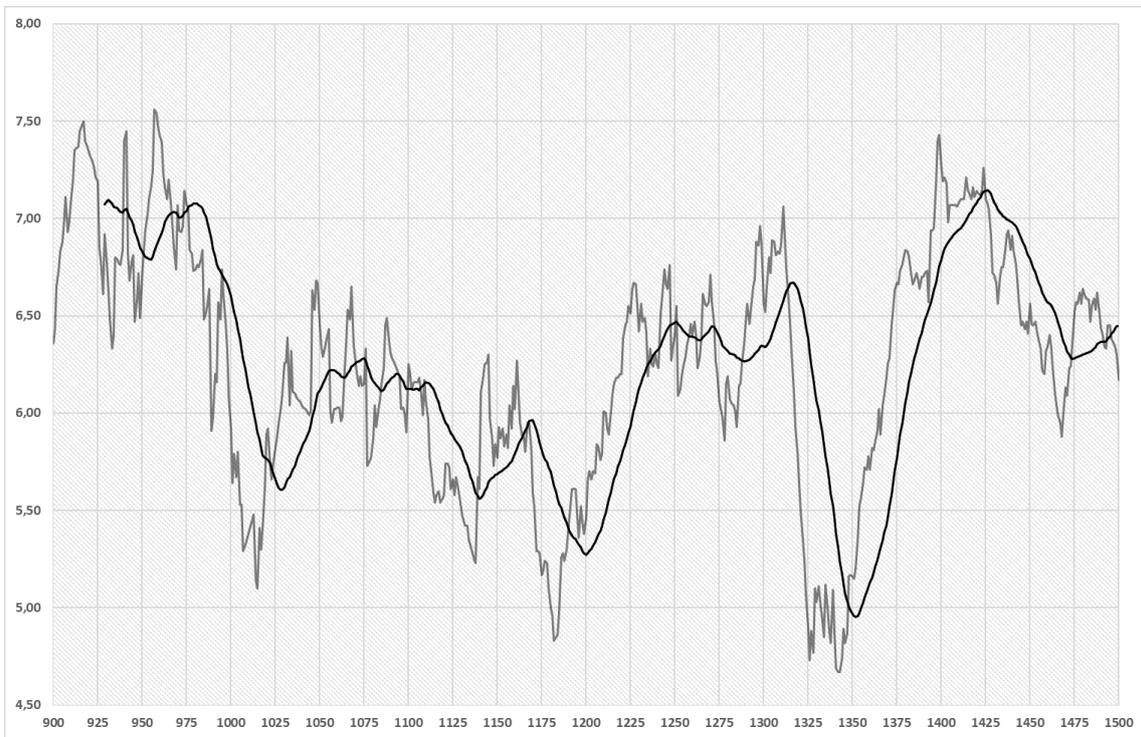

Fig. 4: Reconstructed spring temperatures in Kyōto (Japan) on the basis of the registered start of the cherry blossom, 900-1500 AD (data: Aono/Saito, "Clarifying Springtime Temperature Reconstructions"; dotted line = 30 years average; graph: J. Preiser-Kapeller, OEAW, 2022)

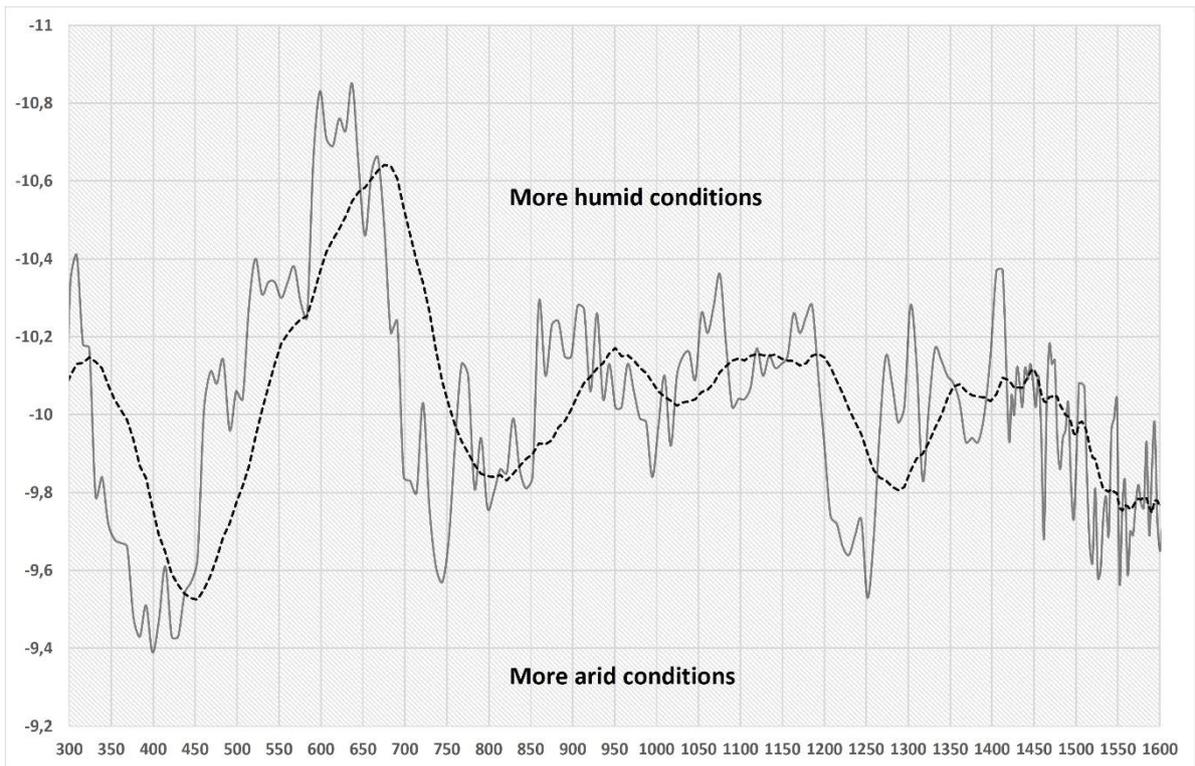

Fig. 5: Sofular Cave (northwestern Turkey) speleothem carbon isotopes record, 300–1600 AD; dotted line = moving average (data: Fleitmann et al., "Sofular Cave"; graph: J. Preiser-Kapeller, OEAW, 2022)